\documentclass[aps,prc,reprint,superscriptaddress,nofootinbib]{revtex4-1}
\usepackage{graphicx}
\usepackage{multirow}
\usepackage{color}
\usepackage{xcolor}

\usepackage{amssymb}
\usepackage{amsmath}
\usepackage{color}
\usepackage{lineno}
\usepackage{enumitem}
\usepackage{comment}
\usepackage[normalem]{ulem}

\def\nuc#1#2{\relax\ifmmode{}^{#1}{\protect\text{#2}}\else${}^{#1}$#2\fi}

\newcommand{\be}{\begin{eqnarray}}
\newcommand{\ee}{\end{eqnarray}}

\newcommand{\bwt}{\begin{widetext}}
\newcommand{\ewt}{\end{widetext}}

\begin{document}

\title{The Lagrange-mesh $R$-matrix method for inhomogenous equations}


\author{Jin Lei}
\email[]{jinlei@pi.infn.it}

\affiliation{Istituto Nazionale di Fisica Nucleare, Sezione di Pisa, Largo Pontecorvo 3, 56127 Pisa, Italy}

\author{Pierre Descouvemont}
\email[]{pdesc@ulb.ac.be}

\affiliation{Physique Nucléaire Théorique et Physique Mathématique, C.P. 229, Université Libre de Bruxelles (ULB), B 1050 Brussels, Belgium}


\begin{abstract}
The Lagrange-mesh $R$-matrix method is generalized to inhomogeneous equations. This method is numerically stable and efficient. It can be directly used for transfer reactions with the formalism discussed by Ascuitto and Glendenning [Phys.~Rev.~181,1396 (1969)] and for inclusive breakup reactions modeled by Ichimura, Austern and Vincent [Phys.~Rev.~C 32, 431 (1985)]. We first present a simple example to assess the method. Then the application to the $^{93}$Nb($d$,$pX$) non-elastic breakup is discussed.
\end{abstract}


\date{\today}%
\maketitle


\section{Introduction}
The $R$-matrix method is a powerful tool in quantum scattering theory. It was first introduced by Wigner and Eisenbud~\cite{Wigner,Wigner2,Wigner3} in the late 1940s in the analysis of resonant nuclear reactions. The resonances were described in terms of compound states formed by the colliding nuclei, and contained in an internal region of the configuration space. 

At present, the main aim of the $R$-matrix theory is to describe scattering states of interacting particles. The configuration space is divided into two regions. The $R$-matrix, which represents the complexity of the compound states, relates the radial component of the wave function to its derivative at the boundary of the internal region. In the external region, it is assumed that the colliding nuclei are weakly interacting, and hence the complexity of the collision process is represented by the $R$-matrix. In early works, the $R$-matrix was represented by a few parameters used to fit experimental data~\cite{RevModPhys.30.257}.

The other aspect of the $R$-matrix theory is that it provides a simple and elegant way of solving the Schr\"{o}dinger equation
\cite{Descouvemont_2010}. It is especially competitive in coupled-channel problems with large numbers of open channels~\cite{DRUET201088}, where the direct integration may become unstable.

On the other hand, most of the scattering problems are traditionally formulated in terms of the transition amplitude. For transfer reactions (single- as well as multi-channel problems), it has been shown by Ascuitto and Glendenning~\cite{Ascuitto} that, instead of using the transition amplitude, one can derive the $S-$matrix from an inhomogeneous equation describing the scattering in the outgoing channels. The inhomogeneity is a source term which describes the production of the residual particle in the transfer process. 

In addition, for the inclusive breakup of two-body projectiles, the nonelastic breakup part in which the participant interacts non-elastically  with the target can be computed by the closed form formula suggested by Ichimura, Austern and Vincent in the 1980s~\cite{IAV85}. The relative wave function in the sub-system is the solution of an inhomogeneous equation. 

Different methods can be used to solve inhomogeneous equations, such as the Green's function~\cite{Glendenning} with Gauss–Legendre quadrature, or the Numerov method~\cite{Ascuitto}. Some applications of the $R$-matrix method
have been performed in atomic physics~\cite{Schneider}. 
There are two important factors to consider when we compare these methods: the efficiency of the solver and the difficulty of obtaining the source terms. 
Normally the $R$-matrix and Green's function methods require less grid points than the Numerov method. This makes the Green's function and $R$-matrix methods more efficient when the source term is complicated.
For example, the $R$-matrix and Green's function methods only require the source term at the quadrature points. However, for the Numerov method, all points of a uniform mesh with a small step are needed. Normally, the number of these points is much larger than the number of quadrature points. Computing the source terms for the $R$-matrix and Green's function methods is therefore much faster than in the Numerov method. 
Another advantage of the $R$-matrix method is the possibility to include non-local interactions.    

Here, we focus on the $R$-matrix method on a Lagrange mesh. Lagrange functions are based on orthogonal polynomials, and
make the calculation of matrix elements very simple \cite{Ba15}. The method has been applied to several problems in atomic as well as in nuclear physics. We extend this formalism to solve inhomogeneous equations and apply it to nonelastic breakup calculations. 

The paper is organized as follows. In Sec.~\ref{sec2}, we present the Lagrange mesh $R$-matrix method for solving the inhomogeneous equations. In Sec.~\ref{sec3}, the formalism is applied to a simple analytical example, and to the nonelastic breakup induced by a deuteron. Finally, we summarize the main results in Sec.~\ref{sec4}. 

\section{Inhomogeneous equations}
\label{sec2}
In this section, we present the Lagrange-mesh $R$-matrix method. In practice, the applications of inhomogeneous equations in nuclear physics are essentially in transfer reactions and in nonelastic breakup reactions. For transfer reactions, the final state is bound, and only a few inhomogeneous equations need to be solved. However, for the nonelastic breakup process, the final states lay on the continuum, and thousands of inhomogeneous equations have to be solved. This means that the Numerov method, which requires a lots of mesh points, is not numerically favorable. 

As the Green's function method is widely used in nonelastic breakup calculations, we present a short outline in the framework of inhomogeneous equations.

\subsection{The $R$-matrix method}
An inhomogeneous Schr\"{o}dinder equation in partial wave $\ell$ is written as
\begin{equation}
\label{eq:inhomo}
\big[ T_\ell(r)+U_\ell(r) -E  \big] u_\ell (r) = \rho_\ell (r),
\end{equation}
with
\begin{equation}
T_\ell(r)=-\frac{\hbar^2}{2\mu}\Big(\frac{d^2}{dr^2}-\frac{\ell(\ell+1)}{r^2}\Big), 
\end{equation}
where $\mu$ is the reduced mass, $U_\ell(r)$ is the effective interaction, $E$ is the center of mass energy and $\rho_\ell(r)$ is the source term. We assume a single-channel problem for the sake of clarity. The extension to multichannel
systems is straightforward.

In the present work, we use the $R-$matrix 
method~\cite{Descouvemont_2010,DESCOUVEMONT2016199,RevModPhys.30.257} to determine the wave functions  
$u_\ell(r)$. The basic idea of the $R-$matrix theory is to divide the space in an internal region 
(with radius $a$) and in an external region. The channel radius $a$ should be large enough so that the
nuclear potential (short range) is negligible. 

For the region outside the channel radius $a$, the potential $U_\ell(r)$ and the source term $\rho_\ell(r)$ tend to zero. The asymptotic part of the radial wave function presents different
forms whether a source term is present or not. With a source term, only outgoing wave are present; we have
\begin{equation}
\label{eq:boundary}
u_\ell^{\text{ext}} (r) =- \mathcal{S}_\ell \mathcal{H}_\ell^+ (\eta,kr), 
\end{equation}
where $\mathcal{S}_\ell$ is the $S-$matrix, and $\mathcal{H}_\ell^+(\eta,kr)$ is an outgoing Coulomb function \cite{Th10} ($k$ is the wave number and $\eta$ is the Sommerfeld parameter). For an homogeneous equation ($\rho_\ell(r)=0$), the external wave function reads
\begin{equation}
u_\ell^{\text{ext}} (r) =\mathcal{H}_\ell^- (\eta,kr)- \mathcal{S}^0_\ell \mathcal{H}_\ell^+ (\eta,kr), 
\end{equation}
where $\mathcal{S}^0_\ell$ is the elastic scattering matrix.

In the internal region ($r \leq a$) the wave function 
is expanded over a set of $N$ basis functions $\varphi_i(r)$ as  
\begin{equation}
\label{eq:wf_expans}
u_\ell^{\text{int}}(r) =\sum_{i=1}^{N} c_{i}^{\ell} \varphi_{i}(r),
\end{equation}
where the choice of function $\varphi_{i}(r)$ will be discussed later. Since these basis functions $\varphi_i(r)$ are valid for $r\leq a$ only, matrix elements of the kinetic energy are not Hermitian. This is addressed by introducing the Bloch operator 
\begin{equation}
\label{eq:bloch}
    \mathcal{L}=\frac{\hbar^{2}}{2 \mu} \delta(r-a)\left(\frac{d}{d r}-\frac{B}{r}\right),
\end{equation}
where $B$ is a boundary parameter, taken here as 
$B=0$. The role of the Bloch operator is twofold: it ensures the hermiticity of the Hamiltonian over 
the internal region, and the continuity of the derivative at the surface. Then, the Bloch-Schr\"{o}dinger equation
equation reads, with a source term
\begin{equation}
\label{eq:bloch_inhomo}
\big[ T_\ell(r)+U_\ell(r) + \mathcal{L} -E  \big] u_\ell^\text{int} (r) = \mathcal{L}u_\ell^\text{ext} (r)+ \rho_\ell (r),
\end{equation}
where $\mathcal{L}u_\ell^\text{ext} (r)$ takes a boundary form which will be discussed later. 

Inserting the expansion (\ref{eq:wf_expans}) into Eq.~(\ref{eq:bloch_inhomo}) provides coefficients $c_i^\ell$ as 
\begin{equation}
\label{eq:coeff}
 c_i^\ell=\sum_j (\boldsymbol{C}^{-1}_{\ell})_{ij}   \big[ \langle \varphi_j |\mathcal{L}| u_\ell^\text{ext}\rangle  + \langle \varphi_j | \rho_\ell\rangle \big],
\end{equation}
where matrix $\boldsymbol{C}_{\ell}$ is given by 
\begin{equation}
\left(\boldsymbol{C}_{\ell}\right)_{i j}=\left\langle\varphi_{i}|T_\ell+U_\ell+\mathcal{L}-E| \varphi_{j}\right\rangle,
\end{equation}
and where $\langle \varphi_j |\mathcal{L}| u_\ell^\text{ext}\rangle$ takes the form 
\begin{equation}
\langle \varphi_j |\mathcal{L}| u_\ell^\text{ext}\rangle = \frac{\hbar^2}{2\mu} \varphi_j(a) \frac{du_\ell^\text{ext}}{dr}. 
\end{equation}

Let us define the $R$-matrix as 
\begin{equation}
\mathcal{R}_\ell = \frac{\hbar^2}{2\mu a } \sum_{ij} \varphi_i(a) (\boldsymbol{C}^{-1}_{\ell})_{ij} \varphi_j(a).
\end{equation}
The continuity condition
\begin{equation}
u_\ell^\text{int} (a)=u_\ell^\text{ext} (a)
\end{equation}
provides the $S$-matrix for the inhomogeneous equation
\begin{equation}
\label{eq:smat}
\mathcal{S}_\ell=\frac{\sum_{ij} (\boldsymbol{C}^{-1}_{\ell})_{ij} \langle \varphi_j | \rho_\ell \rangle \varphi_i(a)}{ka\mathcal{R}_l {\mathcal{H}_\ell^+}' (\eta,ka) - \mathcal{H}_\ell^+ (\eta,ka) },
\end{equation}
where the prime $'$ denotes the derivative with respect to $ka$.
For the homogeneous equation, we get the well known expression of the elastic $S$-matrix
\begin{equation}
\mathcal{S}^0_\ell=\frac{ka\mathcal{R}_l {\mathcal{H}_\ell^-}' (\eta,ka) - \mathcal{H}_\ell^- (\eta,ka)}
{ka\mathcal{R}_l {\mathcal{H}_\ell^+}' (\eta,ka) - \mathcal{H}_\ell^+ (\eta,ka) }.
\end{equation}
The wave function in the internal region is easily determined with coefficients (\ref{eq:coeff}).
Although the $R$-matrix and the Coulomb functions do depend on the channel radius, the $S$-matrices,
as well as the wave functions should not depend on its value, provided it is large enough so that
the nuclear interaction and the source term are negligible. These quantities should be also insensitive to
the number of basis functions $N$. In practice, $N$ is larger when the channel radius increases. The choice
of the channel radius therefore stems from a compromise: it must be large enough to make sure that the
$R$-matrix conditions are satisfied, but as small as possible to reduce the number of basis functions.
The stability of the $S$-matrix is a strong test of the method. As shown in Ref.\ \cite{Descouvemont_2010},
an independent test is also provided by the continuity of the derivative of the wave function at the channel
radius.

\subsection{Lagrange functions}
There are different types of basis functions $\varphi_i(r)$ used in the 
literature~\cite{Descouvemont_2010}. For the numerical simplicity, we choose Lagrange 
functions~\cite{Ba15}, which are defined in the $(0, a)$ interval as
\begin{equation}
\label{eq:lagrange_f}
\varphi_{i}(r)=(-1)^{N+i} \frac{r}{a x_{i}} \sqrt{a x_{i}\left(1-x_{i}\right)} \frac{P_{N}(2 r / a-1)}{r-a x_{i}} , 
\end{equation}
where $P_N(x)$ is the Legendre polynomial of order $N$, and $x_i$ are the zeros of
\begin{equation}
P_{N}\left(2 x_{i}-1\right)=0
\end{equation}
The regularization factor $r/ax_i$ ensures the regular behavior of the basis
functions at the origin. These basis functions satisfy the Lagrange conditions
\begin{equation}
\varphi_{i}\left(a x_{j}\right)=\left(a \lambda_{i}\right)^{-1 / 2} \delta_{i j}, 
\end{equation}
where $\lambda_i$ are the weights of the Gauss–Legendre quadrature
corresponding to the $(0, 1)$ interval. 

If the matrix elements with basis 
functions (\ref{eq:lagrange_f}) are computed at the Gauss approximation of order $N$, consistent 
with the $N$ mesh points, their calculation is strongly simplified. At this approximation, the overlap 
is given by
\begin{equation}
\left\langle\varphi_{i} | \varphi_{j}\right\rangle=\int_{0}^{a} \varphi_{i}(r) \varphi_{j}(r) \mathrm{d} r \approx \delta_{i j}.  
\end{equation}
For a local potential, the matrix elements can be reduced to 
\begin{align}
\langle \varphi_i | U_\ell | \varphi_j  \rangle &= \int_0^a \varphi_i(r) U_\ell(r)\varphi_j(r) dr  \approx  U_\ell(ax_i) \delta_{ij}.
\end{align}
Then the potential matrix elements are given by the values of the potential at the mesh points. This can
be extended to non-local potentials as
\begin{align}
\label{eq:nonlocal}
\langle \varphi_i | U_\ell | \varphi_j  \rangle &= \int_0^a \varphi_i(r) U_\ell(r,r')\varphi_j(r') dr dr'\nonumber \\ & \approx a \sqrt{\lambda_i \lambda_j} U_\ell(ax_i, ax_j).
\end{align}

A matrix element of kinetic energy and Bloch operator, for the case $i=j$,  is given by
\begin{align}
\left\langle\varphi_{i}\right| & T_{\ell}+\mathcal{L}\left|\varphi_{i}\right\rangle \nonumber \\
&=\frac{\hbar^2}{2\mu}\frac{\left(4 N^{2}+4 N+3\right) x_{i}\left(1-x_{i}\right)-6 x_{i}+1}{3 a^{2} x_{i}^{2}\left(1-x_{i}\right)^{2}} \nonumber \\
&+\frac{\hbar^2}{2\mu}\frac{\ell(\ell+1)}{a^2x_i^2}, 
\end{align}
and, for $i\neq j$, by 
\begin{equation}
\begin{aligned}
\left\langle\varphi_{i}\right| & T_{\ell}+\mathcal{L}\left|\varphi_{j}\right\rangle=\frac{\hbar^2}{2\mu}\frac{(-1)^{i+j}}{a^{2}\left[x_{i} x_{j}\left(1-x_{i}\right)\left(1-x_{j}\right)\right]^{1 / 2}} \\
& \times\left[N^{2}+N+1+\frac{x_{i}+x_{j}-2 x_{i} x_{j}}{\left(x_{i}-x_{j}\right)^{2}}\right.\\
&\left.-\frac{1}{1-x_{i}}-\frac{1}{1-x_{j}}\right].
\end{aligned}
\end{equation}
The overlap with the source function, which is needed in the calculation of the $S$-matrix (\ref{eq:smat}), is given by 
\begin{equation}
\langle \varphi_j | \rho_\ell \rangle = \int_0^a \varphi_j(r)\rho_\ell(r) dr \approx \sqrt{a \lambda_j} \rho_\ell(ax_j).
\end{equation}
It should be noted that, by using a Lagrange mesh, the number of basis functions $N$ is also the number of points where the source term needs to be computed. 

\subsection{Green's function method}
The inhomogeneous equation (\ref{eq:inhomo}) can be also solved by the Green's function method with the following integration 
\begin{equation}
\label{eq:greenf}
u_\ell(r) = \frac{2\mu}{\hbar^2 k} \int_0^\infty f_\ell(r_<)h_\ell^{(+)}(r_>) \rho_\ell(r') dr', 
\end{equation}
where $r_<$ stands for min$\{r$, $r'\}$, and
$r_>$ for max$\{r$, $r'\}$. Functions $f_\ell$ and $h_\ell^+$ are the
irregular and regular solutions of the homogeneous equations 
\begin{align}
    \big[ T_\ell(r)+U_\ell(r) -E  \big] f_\ell(r) = 0 ,\nonumber \\
    \big[ T_\ell(r)+U_\ell(r) -E  \big] h_\ell^+(r) = 0 .
\end{align}
The regular solution $f_\ell(r)$ has the same boundary condition as in elastic scattering, whereas $h_\ell^+(r)$ takes the boundary condition,
\begin{equation}
h_\ell^+(kr)  \xrightarrow{r\to\infty} \mathcal{H}_\ell^+ (kr).
\end{equation}
These equations can be solved by the Numerov method. By using the Gauss–Legendre quadrature, Eq.\ (\ref{eq:greenf}) becomes
\begin{equation}
\label{eq:g_psi}
u_\ell(ax_i) \approx \frac{2\mu}{\hbar^2 k} \sum_{j=1}^N f_\ell(ax_<)h_\ell^{(+)}(ax_>) \rho_\ell(ax_j) a\lambda_j, 
\end{equation}
where $x_<$ and $x_>$ stand for min$\{x_i,x_j\}$ and max$\{x_i,x_j\}$, respectively. The $S$-matrix can be obtained by applying Eq.~(\ref{eq:boundary}) at the channel radius. 

One should note that for both the $R$-matrix method and the Green's function method, only a few values of the source term are required. However, for the Numerov method, all the uniform points with a small step size of the source term are needed.

\section{Applications of the $R$-matrix method}
\label{sec3}
In this section, we apply the formalism to a simple, analytical, example and to nonelastic breakup. 
Our goal is to illustrate the theory for different cases and to compare the numerical results with other techniques, such as the Green's function method. The simple example can be easily reproduced by the reader. 
\subsection{Analytical example}
\begin{figure}[tb]
\begin{center}
 {\centering \resizebox*{0.96\columnwidth}{!}{\includegraphics{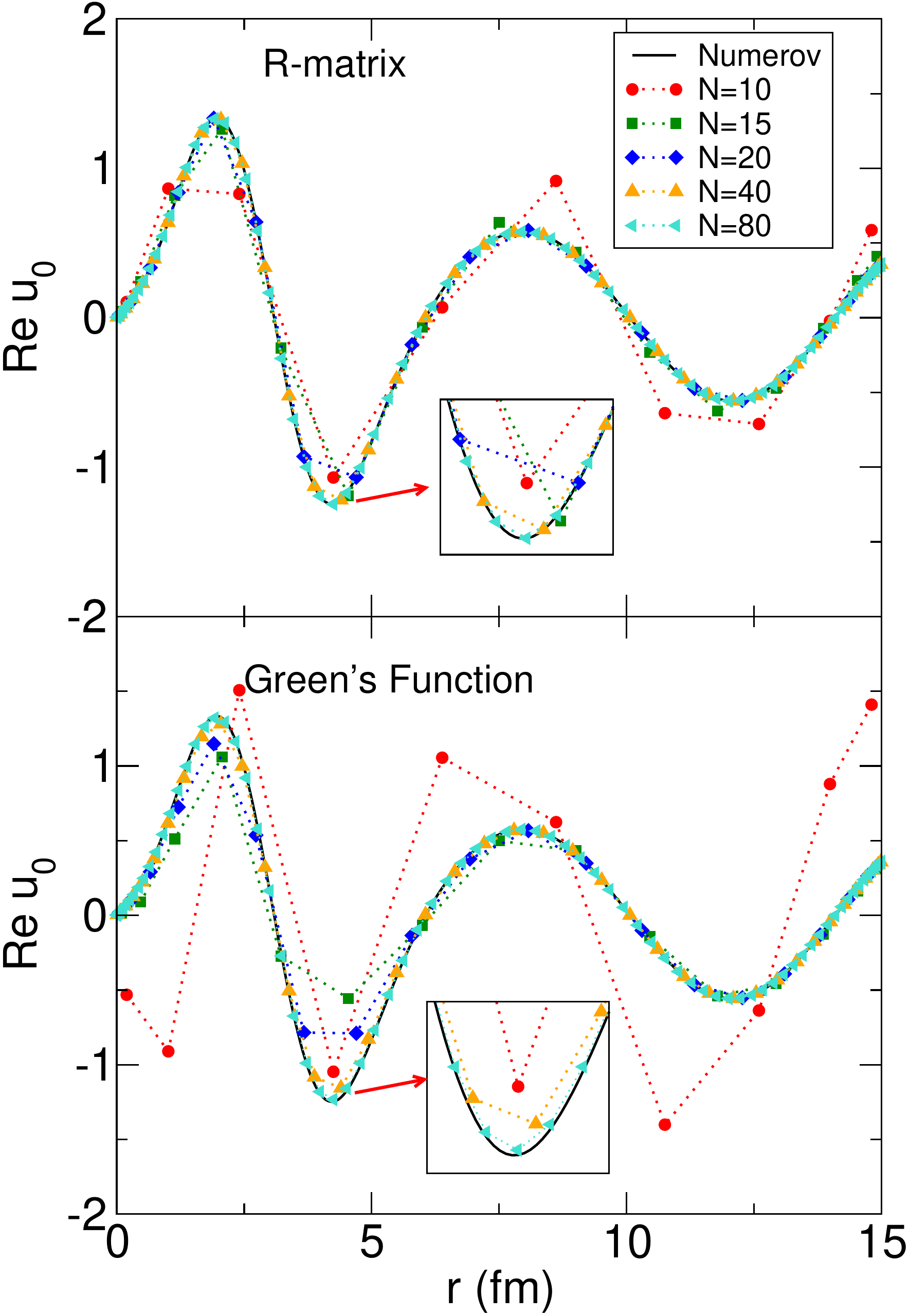}} \par}
\caption{\label{fig:converge}Real part of the wave function $u_0(r)$ for the analytical example of Sec.\ \ref{sec3}.A. The upper and lower panels display the $R$-matrix and Green's function results, respectively.}
\end{center}
\end{figure}
Here we use an analytical example to investigate the $R$-matrix method. We assume that the reduced mass of the system is $\mu=929.4254$ MeV and that the c.m. energy is $E_{cm}=12.74$ MeV. The particles interact through a potential which is local. 
We choose a standard form of the potential, which is defined as
\begin{align}
U(r)=&-V_{r} f\left(r, R_{r}, a_{r}\right) \nonumber \\
&-i W_{v} f\left(r, R_{v}, a_{v}\right)-i W_{s} g\left(r, R_{s}, a_{s}\right),
\end{align}
with 
\begin{equation}
f(r, R, a)=1 \Bigg/\left[1+\exp \left(\frac{r-R}{a}\right)\right],
\end{equation}
and 
\begin{equation}
g\left(r, R, a\right)=-4 a \frac{d}{d r} f\left(r, R, a\right).
\end{equation}
The parameters of the interaction are given by $V_r=77.3$~MeV, $R_r=5.21$~fm, $a_r=0.77$~fm, $W_v=6.1$~MeV, $R_w=6.03$~fm, $a_w=0.47$~fm, $W_s=8.4$~MeV,  $R_s=6.21$~fm, and $a_s=0.77$~fm.  Here we ignore the Coulomb potential. This corresponds to
most  physical applications involving inhomogeneous equations. The inclusion of Coulomb interaction does not affect the final conclusions. 
In our example, we take the source term $\rho_\ell(r)$ as
\begin{equation}
\label{eq:toy_source}
\rho_\ell(r)= U (r)  \sin(r),
\end{equation}
which simulates the shape of realistic source terms. This will be discussed in the next subsection.  

\begin{figure}[tb]
\begin{center}
 {\centering \resizebox*{0.96\columnwidth}{!}{\includegraphics{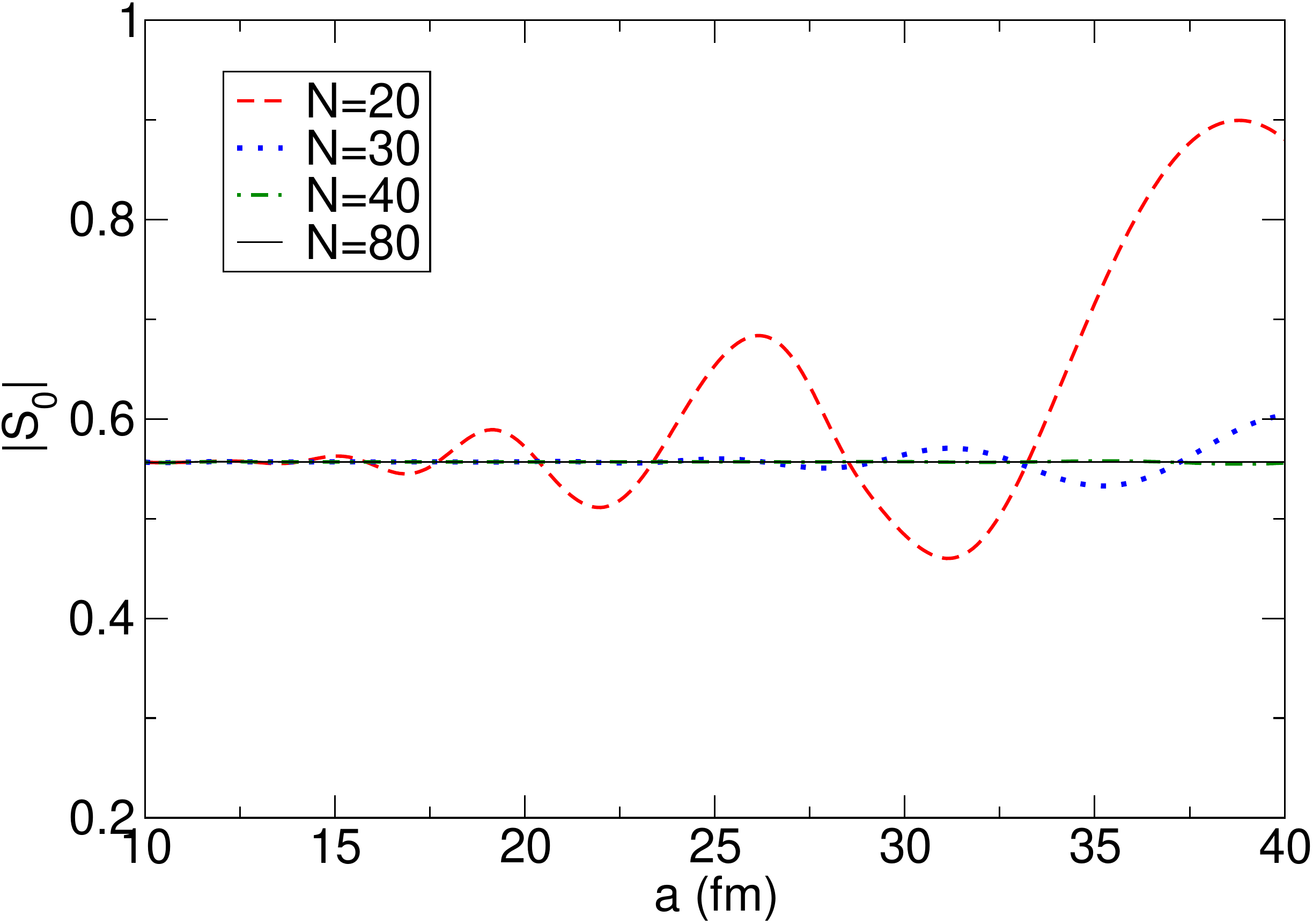}} \par}
\caption{\label{fig:smat}Absolute value of the $s$-wave $S$-matrix  with different channel radii. At the scale
	of the figure, the curves with $N=40$ and $N=80$ are superimposed.  }
\end{center}
\end{figure}

We compare three different methods: the Lagrange-mesh $R$-matrix method, the Green's function method with the Gauss-Legendre quadrature and with the Numerov algorithm to solve this inhomogeneous equation. One should note that when the maximum number of mesh points (quadrature points), $N$, is fixed, the same positions of mesh points are used for both $R$-matrix and Green's function methods. For the Numerov method, a small step uniform mesh (0.05 fm) is used to ensure the convergence.  

In Fig.~\ref{fig:converge}, we show the real part of the $s-$wave solution of the inhomogeneous equation. The channel radius is set at $a=15$~fm. The comparison of the
$R$-matrix method and of the Numerov method is shown in the upper panel. 
It can be found that by increasing $N$, the $R$-matrix method 
agrees very well with the Numerov method. A similar conclusion is drawn 
from the lower panel where the 
Green's function and Numerov methods are compared. 
However, it can be seen that the $R$-matrix method converges 
faster than the Green's function method. With 
a small number of mesh points, $N=20$, the $R$-matrix method provides accurate results. Whereas for the Green's function method, a small number of 
quadrature points can only reproduce the asymptotic region. For the internal part, one has to use a large number of quadrature points (at least $N=80$).

To investigate the numerical properties of the Lagrange mesh $R$-matrix method, we show the absolute value of the $S$-matrix for the $s-$wave. It is computed with different $N$ values and channel radii in Fig.~\ref{fig:smat}. As expected, small values of the channel radius $a$ require small bases. For example, for $a\approx 10$ fm, $N=20$ fairly reproduces the correct $S$-matrix,  whereas, for $a=20$ fm, at least $N=30$ is required.

\begin{figure}[tb]
\begin{center}
 {\centering \resizebox*{0.96\columnwidth}{!}{\includegraphics{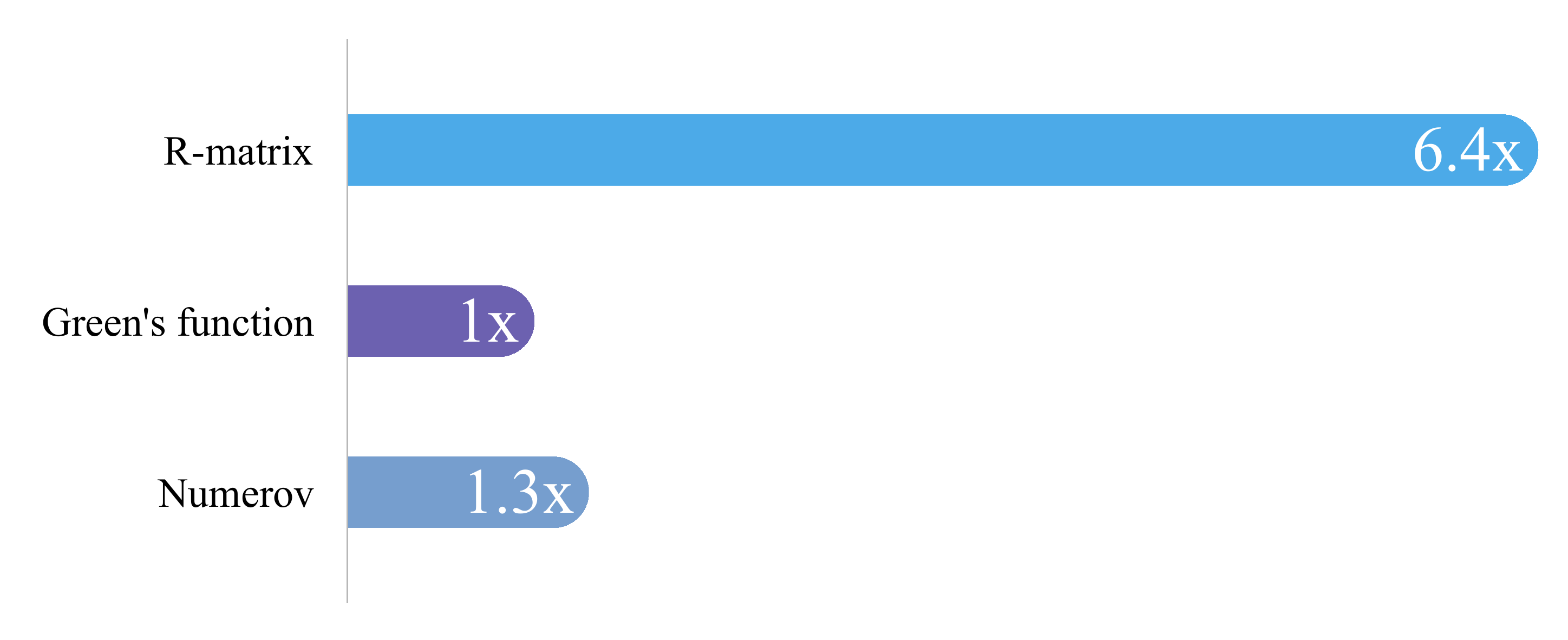}} \par}
\caption{\label{fig:time}Comparison the efficiency of the $R$-matrix, Green's function, and Numerov methods, taking the Green's function method as unit.}
\end{center}
\end{figure}

We also compare the efficiency of the three methods. For that, we consider a large number of inhomogeneous equations, and measured the CPU time with the current implementation~\cite{inhomoR}. The results are shown in Fig.~\ref{fig:time}, in which we take the Green's function as unit. It can be seen that, the $R$-matrix method is the fastest one which is about 6 times faster than the Green's function method. The Green's function is the slowest one, since one has to use Numerov method to obtain the regular and irregular parts of the Green's function. The testing code can be found in Ref.~\cite{inhomoR}.



\subsection{$^{93}$Nb($d$,$pX$) nonelastic breakup}
In the second example, we consider the inclusive breakup reaction of deuterons on a $^{93}$Nb target in which only the outgoing proton is detected. This 
reaction was analyzed in detail in Ref.~\cite{Jin15,Jin19}. We can schematically write it as
\begin{equation}
d+^{93}{\rm Nb}\to p + (^{93}\text{Nb} +n)^*,  
\end{equation}
where notation $()^*$ denotes any possible state of 
the $^{93}$Nb+n system. This includes the elastic 
breakup (EBU) process, in which both $p$ and $n$ 
scatter elastically from $^{93}$Nb, and hence
the latter is left in its ground state. The other 
contributors, which we call globally non-elastic 
breakup (NEB), are those in which $n$ undergoes a non-elastic interaction with the target, including $n$ + $^{93}$Nb inelastic scattering and fusion. 

Here we focus on solving the NEB part with the $R$-matrix method. By using the three-body model proposed by Ichimura, Austern and Vincent (IAV) \cite{IAV85}, the NEB cross section is given by the closed-form formula
\begin{equation}
\left.\frac{d^{2} \sigma}{d E_{p} d \Omega_{p}}\right|_{\mathrm{NEB}}=-\frac{2}{\hbar v_{d}} \rho_{p}\left(E_{p}\right)\left\langle\varphi_{n}(\vec{k}_{p})\left|\operatorname{Im}\left[U_{n}\right]\right| \varphi_{n}(\vec{k}_{p})\right\rangle.
\end{equation}
In this definition, $\rho_p(E_p)$ is the proton density of states, $v_d$ is the velocity of the deuteron, 
$U_n$ is an optical potential describing the $n$ + $^{93}$Nb elastic scattering, and
$\varphi_{n}$($\vec{k}_p$, $\vec{r}_n$) is a relative wave 
function describing the motion between $n$ and $^{93}$Nb when a  proton is scattered with momentum $\vec{k}_p$. This function is obtained by solving the inhomogeneous equation
\begin{equation}
\label{eq:iavpsi}
(E_n-T_n-U_n)\varphi_n(\vec{k}_p,\vec{r}_n) = 
\langle \vec{r}_n \chi_p^{(-)}  |V_\text{post}|\Psi^{3b(+)}\rangle,
\end{equation}
where $E_n=E^{3b}-E_p$ and $T_n$ are the energy and kinetic energy in the $n$-$^{93}$Nb subsystem respectively, and $E^{3b}$ is the three-body energy in the center of mass frame. In this definition, $\chi_p^{(-)*}(\vec{k}_p, \vec{r}_p)$ is the distorted wave describing the relative motion between $p$ and the $n+^{93}$Nb compound system (obtained with some optical potential $U_p$), $V_\text{post}$ is the post-form transition operator and $\Psi^{3b(+)}$ is the three-body scattering wave function. It has been found that the DWBA wave function is a good approximation for the three-body wave function~\cite{Jin19}. Therefore we take
\begin{equation}
 \Psi^{3b(+)}\approx \Psi^{\mathrm{DWBA}(+)}= \chi_d\phi_d, 
\end{equation}
where $\chi_d$ is the distorted wave describing the relative motion between the projectile and the target, and $\phi_d$ is the bound-state wave function of deuteron. 
The partial-wave expansion of the above equations for nonelastic breakup can be found in Refs.~\cite{Jin15,Jin18}. We adopt the same potentials.

\begin{figure}[tb]
\begin{center}
 {\centering \resizebox*{0.96\columnwidth}{!}{\includegraphics{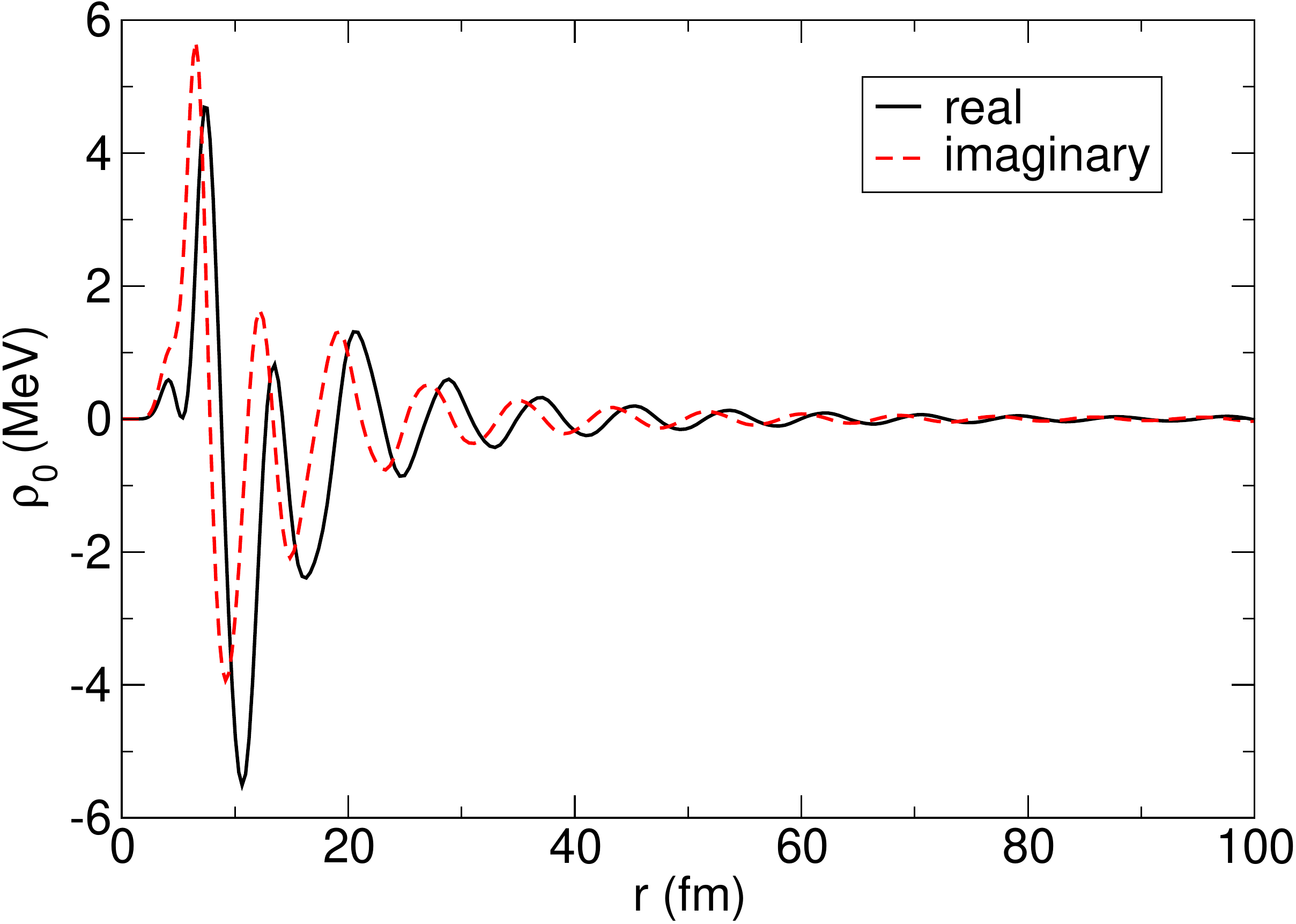}} \par}
\caption{\label{fig:source}Real and imaginary parts of the source term function in the $^{93}$Nb($d$,$pX$) reaction at $E_{lab}=25.5$ MeV for a outgoing proton energy of 14 MeV, and for the partial wave set of $\ell_d=8$, $\ell_p=6$, and $\ell_n=8$. }
\end{center}
\end{figure}

\begin{figure}[tb]
\begin{center}
 {\centering \resizebox*{0.96\columnwidth}{!}{\includegraphics{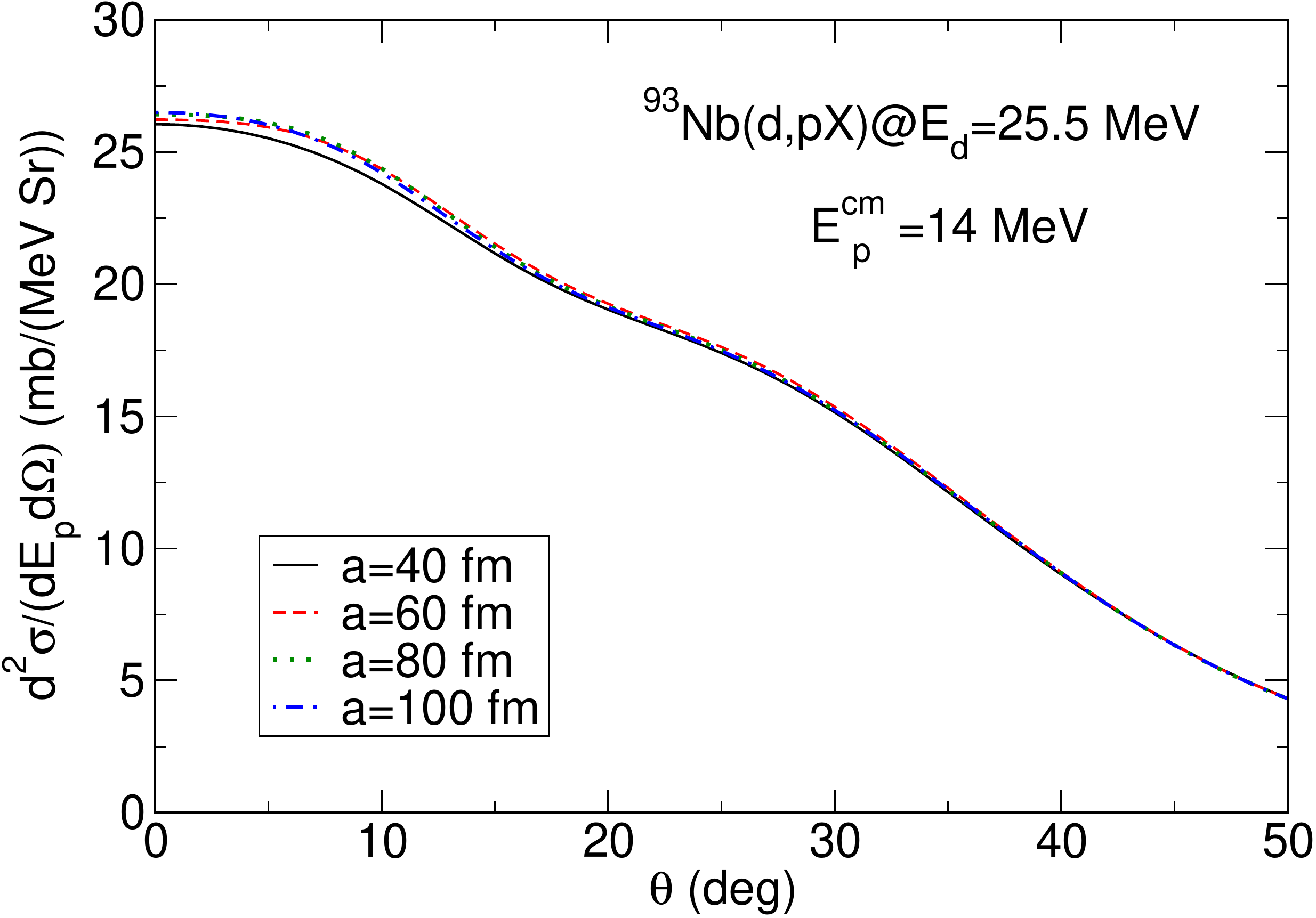}} \par}
\caption{\label{fig:dsdw_diff_a} Sensitivity to the channel radius of the NEB double differential $^{93}$Nb($d$,$pX$) cross section at $E_{lab}=25.5$ MeV for a outgoing proton energy of 14~MeV.}
\end{center}
\end{figure}

\begin{figure}[tb]
\begin{center}
 {\centering \resizebox*{0.96\columnwidth}{!}{\includegraphics{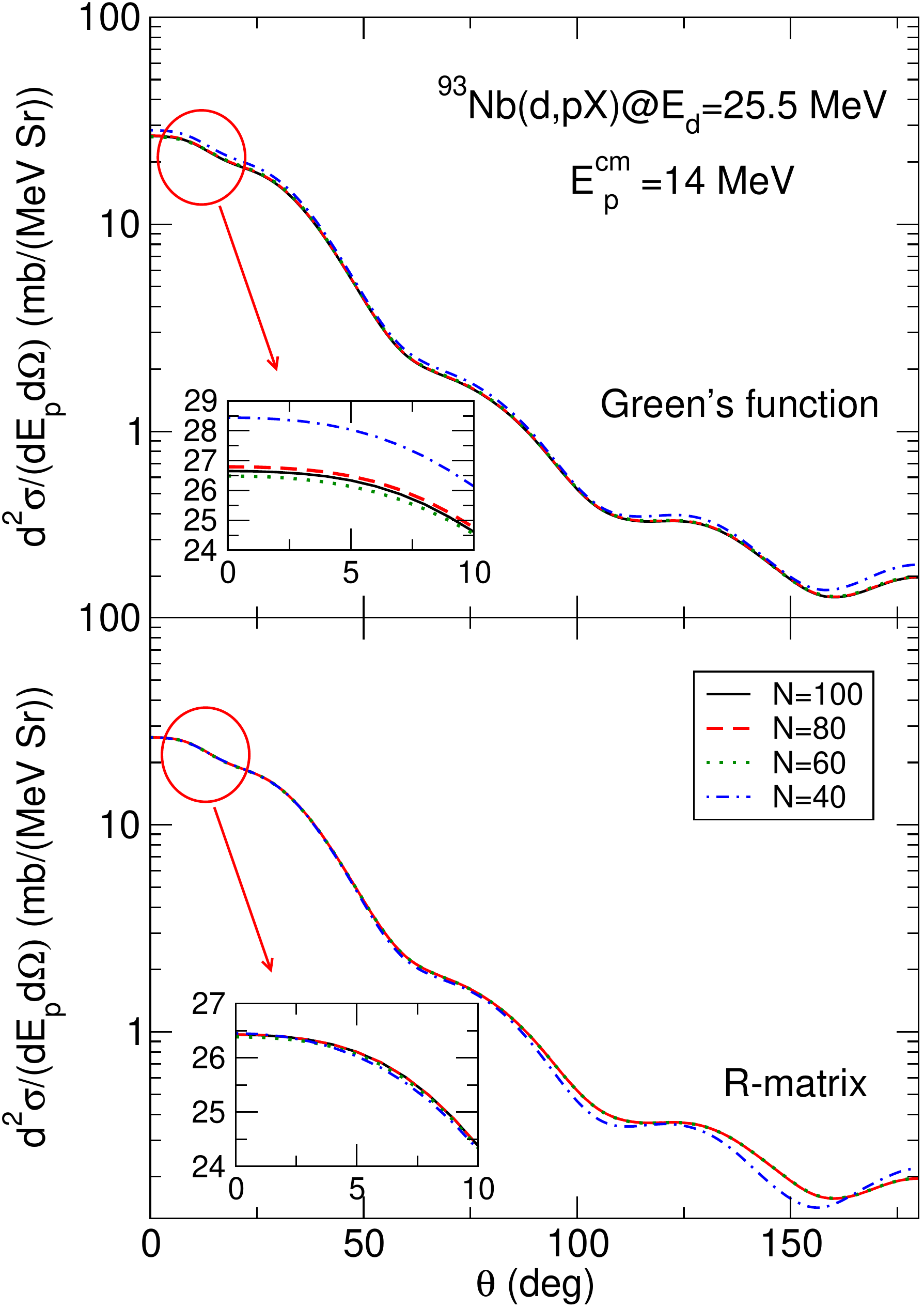}} \par}
\caption{\label{fig:dsdedw_converge}Convergence of the nonelastic breakup double differential cross section of the $^{93}$Nb($d$,$pX$) reaction at $E_{lab}=25.5$ MeV for a outgoing proton energy of 14~MeV. The calculations are done with the Green's function method (upper panel) and with the $R$-matrix method (lower panel).}
\end{center}
\end{figure}

We employ the Green's function and $R$-matrix 
methods to solve the inhomogeneous 
equation~(\ref{eq:iavpsi}) in its equivalent prior form.
The relation between its post and prior forms can be found in Refs.~\cite{Jin15b,Jin18b}.
In Fig.~\ref{fig:source}, we show an example of the source term for the partial waves $\ell_d=8$, $\ell_p=6$, and $\ell_n=8$ calculated by the prior form IAV model.  It can be seen that the source term function starts from zero, then oscillates, and finally tends to zero again. This justifies the choice made in the analytical example (\ref{eq:toy_source}).

In addition, we note that this source term presents a long range compared to the nuclear potential. 
A large channel radius is therefore needed in the $R-$matrix calculation.
To verify this point, we show the comparison of NEB double-differential cross cross sections computed by different channel radii in Fig.~\ref{fig:dsdw_diff_a}. It can be seen that there are some differences at small angles between $a<80$ fm and $a\geq 80$ fm. This shows that, to have a high accuracy at small angles, the long-range source term needs a large channel radius.

In Fig.~\ref{fig:dsdedw_converge}, we show a convergence test for the same reaction. The calculations are done with a channel radius $a=80$~fm, where the calculated cross sections are converged. A clear difference between $N=40$ and $N>40$ can be found for both methods. 
In general, about $(3-5)$ mesh points are needed for each interval of length $\pi/k_n$, where $k_n$ is the wave number of $n-^{93}$Nb subsystem. Then, the minimum mesh number required by the $R$-matrix method for a given channel radius $a$ can be estimated by using the following relation: $N\approx (3-5)ak_n/\pi$. In the present case, we have $E_n =8.7$ MeV, and $\pi/k_n\sim 5$ fm. The the simple relation gives $N\approx48-80$ for $a=80$ fm.
On the other hand, the $R$-matrix method converges much faster
than the Green's function method, one can not see any different of the cross sections when $N \geq 60$. 
As we found in the analytical example, the convergence of the Green's function method is slow. 

It should also be noted that the source term 
$\rho(\vec{r}_n)=\langle \vec{r}_n\chi_p^{(-)}  |V_\text{post}|\Psi^{3b(+)}\rangle$ is the 
time-consuming part in the numerical calculations 
using the partial wave method. For each value of $r_n$, one has to perform a
transformation from the incoming Jacobi coordinates, $(n+p)+^{93}$Nb, to the 
outgoing Jacobi coordinates, $(n+^{93}$Nb)+$p$.
In practice, this makes the Numerov method time consuming, since it requires many grid points. For the current application, 1600 points are needed by using a step size of $0.05$~fm, compared to 60 points used in the $R$-matrix method. 
In addition, when the effective interaction $U_n$ is non-local it is more natural to use the $R$-matrix method, since the matrix elements of a non-local potential are trivial [see Eq.~(\ref{eq:nonlocal})].

\section{Summary}
\label{sec4}
In summary we have addressed the problem of solving inhomogeneous equations with the Lagrange-mesh $R$-matrix method. For that purpose, we derived the Lagrange-mesh $R$-matrix formulas for inhomogeneous equations and applied them to solve an analytical example and compared the solutions with Green's function and Numerov methods. After that, we also applied the formalism to the NEB of a deuteron induced reaction. 
Our study shows that the Lagrange-mesh $R$-matrix method is a fast and accurate technique for solving inhomogeneous equations.

To compare the solution of the different methods, there are two factors that need to be considered, the efficiency of the solver and the difficulty of obtaining the source term. The $R$-matrix is the most efficient tool regarding both aspects. The present method can be easily extended to multi-channel problems. Also, calculations involving large bases can be made faster by using propagation techniques (see for example Ref.~\cite{DESCOUVEMONT2016199} and references therein).

\bigskip
\begin{acknowledgments}
The authors are grateful to Antonio M. Moro and Angela Bonaccorso for a critical reading of the manuscript and helpful discussions.
This work was supported by the Fonds de la Recherche Scientifique - FNRS under Grant Numbers 4.45.10.08 and J.0049.19.
\end{acknowledgments}

\bibliography{inclusive_prc}
\end{document}